\documentclass[aps, prd, preprint, groupedaddress,showpacs,18pt]{revtex4}
\usepackage{hyperref}
\usepackage{mathrsfs}
\usepackage{bbm}
\usepackage{amsfonts}
\usepackage{amsmath}
\usepackage{graphicx}
\usepackage{subfig}
\usepackage{feynmp}

\newcommand{\void}[1]{}

\def\be{\begin{equation}}
\def\ee{\end{equation}}
\def\bea{\begin{eqnarray}}
\def\eea{\end{eqnarray}}
\def\lmd{\lambda}
\def\la{\langle}
\def\ra{\rangle}

\newcommand{\td}[1]{\tilde{#1}}

\newcommand{\bary}{\begin{array}}
\newcommand{\eary}{\end{array}}

\def\nb{\nonumber}

\begin{document}

    \title{Complexified Ward Identity in pure Yang-Mills theory at tree-level}

\author{
Gang~Chen,$^{1}$
}
{\affiliation{
$^{1}$Department of Physics, Nanjing University\\
22 Hankou Road, Nanjing 210098, China
}

\hspace{1cm}
\begin{abstract}
Up until now, the BCFW technique has been a widely used method in getting the amplitudes in various theories. Usually, the vanishing of the boundary term is necessary for the efficiency of the method. However, there are also many kinds of amplitudes which will have boundary terms. Hence it will be nice if it is possible to get the boundary terms in an efficient manner. As is well-known, in gauge theory the Ward identity imposes constraints  on the form of the amplitude. In \cite{Chen},  we studied the Ward identity with a pair of shifted lines and the implied recursion relations. In this article, we discuss the complexified Ward identity in more detail. In particular we give a proof of the complexified Ward identity directly from the Feynman rules in Feynman-Lorentz gauge. Furthermore, we give more examples in calculating the one-line off-shell currents using this technique.  
\end{abstract}


\date{\today}
\maketitle
\section{Introduction}
In the early development of Quantum mechanics, A. Einstein had proposed the possibility of constructing the Quantum theory from a purely algebraic theory.  At that time, nobody knows how to obtain the basis of such a theory. In Quantum Field Theory (QFT), it seems even more difficult for this aim except for some special case such as Topological field theory. Recent years, there are exiting progress (BCFW) on the amplitudes in QFT, which is helpful for constructing an QFT from algebra system. At tree level, the amplitudes in pure Yang-Mills theory can be taken as rational functions of external momenta and external states in spinor forms 
\cite{Parke:1986gb,Xu:1986xb,Berends:1987me,Kosower,Dixon1,Witten1}.  Such rational functions are probable to study in detail in algebra system now. According to this, BCFW recursion relation was proposed and developed in \cite{Britto:2004nj,Britto:2004nc,Britto:2004ap}. It was then proven in \cite{Britto:2005fq} using the singularity properties of the tree-level on-shell amplitudes.  For the theory with  massive fields \cite{Badger1,Ozeren,Schwinn,Chen1,Chen2}, the amplitudes are also rational functions of external momenta and states in spinor forms.     

At loop level,  although the total amplitudes are not rational functions in general any more, they can be decomposed in some basic scalar integrals with coefficients being rational functions of external spinors \cite{BernD1,BernD2}. The coefficients structures are also studied deeply in \cite{Dixon4,Bern,Bern1}. At loop level, the integrands of the amplitudes are rational functions of the external spinors and integral momenta. For the N=4 planar super Yang-Mills theory, \cite{Nima}  give an explicit recursive formula for the all-loop integrand for scattering amplitudes.

In gauge theory, Ward identity is a constraint for the amplitudes at any loop-level. Hence it lead a algebra constraint for the rational parts in the amplitudes. In this article, we discuss more about Ward identity. We complexify the momenta of a pair of external lines in a proper way and give a proof of the Ward identity with complex external momenta using Feynman rules. In BCFW formalism, the vanishing of the boundary term is necessary for the application of the recursion relation. However, there are still various kinds of amplitudes which do have boundary terms. In this article, we will use the complexified Ward identity  to determine the amplitudes in gauge theory.  For the poles at finite position, the residues are the same as those in BCFW recursion relations.  Moreover, this Ward identity will lead to  new forms for the boundary terms, which can be obtained by another recursion relation.  We will focus on the vector off-shell current which is the amplitude with one external line amputated and its momenta extended to off-shell. The method can also be extended to the tensor off-shell currents with several external off-shell lines. Our method is particularly useful for the case that the on-shell lines do not have the same helicity structures. In this sense, our technique is complementary to the off-shell current recursion relation presented in \cite{Berends:1987me,Kosower}.

\section{Complexified Ward Identity}
In an  interacting gauge theory,  the Ward identity is a consequence  of the gauge current conservation .  As a result, the tensor currents  vanish when contracted with all the momenta of the external off-shell lines
\be
k^1_{\mu_1} k^2_{\mu_2}\cdots k^m_{\mu_m} \mathcal{A}^{\mu_1\mu_2\cdots \mu_m}=0,
\ee  
where $m$ is the number of the external off-shell lines and $k^i_{\mu_i}$ is the momenta of the external off-shell  lines.. The tensor current $\mathcal{A}^{\mu_1\mu_2\cdots \mu_m}$ is defined to be the amplitude with removing the propagators of the external off shell lines, while in \cite{Dixon1}, they keep the propagators of the external off-shell lines in the definition.   This subtle difference will lead the current formalisms in this article seem to be different with those in  \cite{Feng}. However  actually  our results are identical with those in  \cite{Feng}.   

As long as the current conservation is not broken by  quantum corrections, the identity holds at each level of the perturbative expansion.  Furthermore, at tree level, the Ward identity holds even when the momenta are complexified with the on-shell condition untouched.   Hence it contains much more information about the off-shell currents than just the current conservation.  There is a simple proof for the complexified Ward identity directly according to the Feynman rules in Lorentz-Feynman gauge, where the tree level 2, 3 and 4 point vertices take the forms
\bea
V^2_{\mu\nu}&=&-i{\eta_{\mu\nu}\over k^2}\nb\\
V^3_{\mu_1\mu_2\mu_3}&=&{i\over\sqrt 2}\left(\eta_{\mu_1\mu_2}(k_1-k_2)_{\mu_3}+\eta_{\mu_2\mu_3}(k_2-k_3)_{\mu_1}+\eta_{\mu_3\mu_1}(k_3-k_1)_{\mu_2}\right)\nb\\
V^4_{\mu_1\mu_2\mu_3\mu_4}&=&{i\over 2}\left(2\eta_{\mu_1\mu_3}\eta_{\mu_2\mu_4}-\eta_{\mu_1\mu_2}\eta_{\mu_3\mu_4}-\eta_{\mu_1\mu_4}\eta_{\mu_2\mu_3}\right).
\eea

We prove the complexified Ward identity by induction. Firstly, it is easy to check directly that the complexified Ward identity still holds for the currents with two and three on-shell lines. Secondly, we assume the complexified Ward identity holds for the currents with the external on-shell lines less than N.  Then we need to prove the currents with N on-shell lines are conserved.  

To this end, we first the classify the diagrams according to the types of the vertex $\bar V$ connecting directly to the external off-shell line and of the nearest two vertices $V_a$ and $V_b$ of on the left and right hand the off-shell line. Hence in total, there are five kinds of diagrams. 

Case 1: $\bar V$ is 4-gluon vertex. Then we denote the result of the current contracted with the momentum of external off-shell line as $F^N(n_1, n_2, n_3)$, where $n_1, n_2, n_3$ are the positive numbers of external lines in $J_1, J_2, J_3$ respectively. And $n_1+n_2 +n_3=N$. Similarly for the other four cases;

Case 2: $\bar V$, $V_a$ and $V_b$ are 3-gluon vertices. Then we denote the result as $T^N(n_1, n_2 \vdots m_1, m_2)$. The v-dots denotes the position of the off-shell external line;

Case 3: $\bar V$, $V_a$ are 3-gluon vertices and  and $V_b$ is 4-gluon vertex. Then we denote the results as $T^N(n_1, n_2 \vdots m_1, m_2, m_3)$;

Case 4: $\bar V$, $V_b$ are 3-gluon vertices and  and $V_a$ is 4-gluon vertex. Then we denote the results as $T^N(n_1, n_2 ,n_3\vdots m_1, m_2)$;

Case 5: $\bar V$ is 3-gluon vertex and  and $V_a$, $V_b$ are 4-gluon vertices. Then we denote the results as $T^N(n_1, n_2, n_3 \vdots m_1, m_2, m_3)$. 

Then according to inductive assumption about the current conservation for fewer external on-shell lines,  when multiplied with the external off-shell momentum,  in case 2-5,  $\bar V^3_{\mu_1\mu_2\mu_3}k_3^{\mu_3}={i\over\sqrt 2}\eta_{\mu_1\mu_2}(k^2_2-k^2_1)$.  We divide it into $\bar V^3_{\mu_1\mu_2\mu_3}k_3^{\mu_3}=V^R_{\mu_1\mu_2}k^2_2-V^L_{\mu_1\mu_2}k^2_1\equiv V^R_{\mu_1\mu_2}+V^{-L}_{\mu_1\mu_2}$. Correspondingly, $T^N=T^N_{R}+T^N_{-L}$.

To make the cancelation obvious, we arrange the terms for all the diagrams as follows
\bea\label{Allt}
&&T^N(1 \vdots 1, N-2)_{R} +T^N(1 \vdots 2, N-3)_{R}+\cdots + T^N(1 \vdots N-2, 1)_{R}\nb\\
&+&T^N(1 \vdots 1, 1, N-3)_{R}+T^N(1 \vdots 1, 2, N-4)_{R}+\cdots + T^N(1 \vdots N-3, 1, 1)_{R}\nb\\ \hline\nb\\
&+&T^N(1,1 \vdots 1, N-3)_{R-L} +T^N(1,1 \vdots 2, N-4)_{R-L}+\cdots + T^N(1,1 \vdots N-3, 1)_{R-L}\nb\\
&+&T^N(1,1 \vdots 1, 1, N-4)_{R-L}+T^N(1,1 \vdots 1, 2, N-5)_{R-L}+\cdots + T^N(1,1 \vdots N-4, 1, 1)_{R-L}\nb\\
\hline \nb\\
&+&T^N(1,2 \vdots 1, N-4)_{R-L} +T^N(1,2 \vdots 2, N-5)_{R-L}+\cdots + T^N(1,2 \vdots N-4, 1)_{R-L}\nb\\
&+&T^N(1,2 \vdots 1, 1, N-5)_{R-L}+T^N(1,2 \vdots 1, 2, N-6)_{R-L}+\cdots + T^N(1,2 \vdots N-5, 1, 1)_{R-L}\nb\\
&&\cdots\cdots\cdots\cdots\cdots\cdots\cdots\cdots\cdots\cdots\cdots\cdots\cdots\cdots\cdots\cdots\cdots\cdots\cdots\cdots\cdots\cdots\cdots\nb\\
&+&T^N(1,1,1 \vdots 1, N-4)_{R-L} +T^N(1,1,1 \vdots 2, N-5)_{R-L}+\cdots + T^N(1,1,1 \vdots N-4, 1)_{R-L}\nb\\
&+&T^N(1,1,1 \vdots 1, 1, N-5)_{R-L}+T^N(1,1,1 \vdots 1, 2, N-6)_{R-L}+\cdots + T^N(1,1,1 \vdots N-5, 1, 1)_{R-L}\nb\\
&&\cdots\cdots\cdots\cdots\cdots\cdots\cdots\cdots\cdots\cdots\cdots\cdots\cdots\cdots\cdots\cdots\cdots\cdots\cdots\cdots\cdots\cdots\cdots\nb\\
&+&T^N(2,1 \vdots 1, N-4)_{R-L} +T^N(2,1 \vdots 2, N-5)_{R-L}+\cdots + T^N(2,1 \vdots N-4, 1)_{R-L}\nb\\
&+&T^N(2,1 \vdots 1, 1, N-5)_{R-L}+T^N(2,1 \vdots 1, 2, N-6)_{R-L}+\cdots + T^N(2,1 \vdots N-5, 1, 1)_{R-L}\nb\\
\hline\nb\\
&\vdots&\nb\\
&\vdots&\nb\\
\hline \nb\\
&+&T^N(N-2,1 \vdots 1)_{-L}  \nb\\
&\vdots&\nb\\
&+&T^N(1,N-2 \vdots 1)_{-L} \nb\\
&+&T^N(1,1,N-3 \vdots 1)_{-L}  \nb\\
&+&T^N(1,2,N-4 \vdots 1)_{-L}  \nb\\
&\vdots&\nb\\
&+&T^N(N-3,1,1 \vdots 2,1)_{-L} \nb\\
\hline \nb\\
&+&\sum_{n_1,n_2,n_3} F^N(n_1,n_2, n_3)
\eea
Under the recursive assumption, we can verify directly that the summations of diagrams as shown in Fig.1  and Fig.2 will vanish.  That is 
\bea\label{twocan}
&&T(\{n_1\}\vdots n_2,n_3)_R+T(n_1,n_2\vdots \{n_3\})_{-L}+F(n_1,n_2,n_3)=0 \nb\\
&&T(\{n_1\}\vdots n_2, n_3, n_4)_R+T(n_1,n_2,n_3\vdots \{n_4\})_{-L}=0,
\eea
where $\{n\}$ denotes all the possible divisions of $n$ into $(i,j)$ and $(i,j,k)$. 
Then we can find, in (\ref{Allt}), in each box  between two horizontal lines, the summation of the $L$-terms with fixed pairs $(n_1, n_2)$ and triples $(m_1, m_2, m_3)$ to the left of the off-shell line are $T(n_1,n_2\vdots \{n_3\})_L$ and  $T(m_1,m_2,m_3\vdots \{m_4\})_L$ respectively.  And  similarly the summation of the $R$-terms with fixed pairs $(i_1,i_2)$ and triples $(j_1,j_2,j_3)$ to the right of the off-shell line  are $T(\{i_3\}\vdots i_1,i_2)_R$ and $T(\{j_4\}\vdots j_1, j_2, j_3)_R$.  Then we can arrange all the terms into the groups as in (\ref{twocan}). Hence  they cancel exactly with each other. 

\begin{figure}[htb]
\centering
\includegraphics{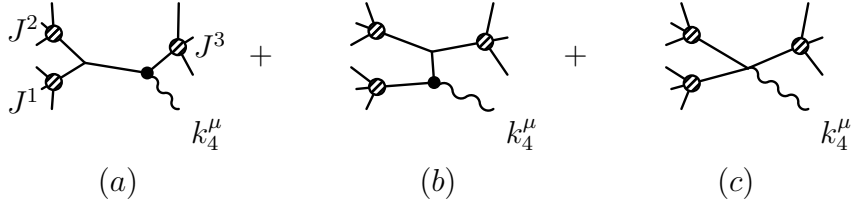}
\caption{In thees diagrams we use $\bullet$  to denote the $\bar V_{-L}$ and  $\bar V_{R}$ to (a) and (b) respectively.}
\end{figure}

\begin{figure}[htb]
\centering
\includegraphics{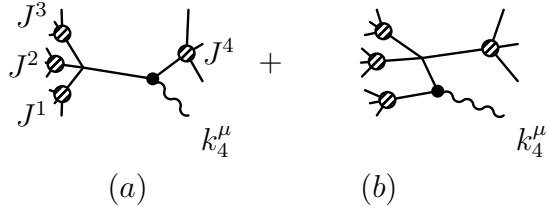}
\caption{Similarly as Fig.1, the $\bullet$  denote the $\bar V_{-L}$ and  $\bar V_{R}$ for (a) and (b) respectively.}
\end{figure}

Now we shift the momenta of a pair of lines $i,v$, where $i$ is one on-shell line with external state $\epsilon_i$ and $v$ is the off-shell line.  The momentum shift is chosen such as to keep the momentum conserved and the $i$-line on-shell.  For example,  we can shift the momenta as $\hat p_i=p_i-z\eta_i$ and  $\hat p_v=p_v-z\eta_i$, where $z$ is an arbitrary complex parameter and $\eta_i$ is a four-vector which  satisfies $\eta_i\cdot \epsilon_i=0$. Then we get a complexified form of the Ward identity.  Acting with the first order derivative  with respect to $z$ on the complexified Ward identity, we obtain
\be\label{recz}
\mathcal{\hat A}(z)_\mu \eta^\mu_i=-\hat p_v^\mu {d\mathcal{\hat A}(z)_\mu\over dz},
\ee
where $\mathcal{\hat A}(z)_\mu$ denotes the complexified vector off-shell current.

If the  shifted on-shell line is in helicity state $\epsilon_i^+={\mu\td\lmd_i\over \langle\mu,\lmd_i\rangle}$,   we shift the momenta as $\lmd_i\rightarrow\lmd_i-z\lmd_v, \td\lmd_v\rightarrow\td\lmd_v+z\td\lmd_i$, and $\eta_i=\lmd_v\td\lmd_i$.  It is easy to see that such momenta shift only lead to a $z$-dependent in the  denominator of $\epsilon_i^+$.   To avoid the unphysical pole in the  denominator of $\epsilon_i^+$, we  choose the reference spinor as $\mu=\lmd_v$.   Then  $\epsilon_i^+$ do not dependent on $z$ even under the momenta shift and we do not get a extra contribution from the external wave function $\epsilon_i^+$ in the left hand of (\ref{recz}). Hence the choice of the momenta shift and the reference spinor above are convenient for the calculation in practice. Similarly, for the negative state $\epsilon_i^-={\lmd_i\td\mu\over \langle\td\lmd_i,\td\mu\rangle}$, we can shift its  momentum  as $\td\lmd_i\rightarrow\td\lmd_i-z\td\lmd_v, \lmd_v\rightarrow\lmd_v+z\lmd_i$. For the same reason, the reference spinor is taken as $\td\mu=\td\lmd_v$. 

For pure Yang-Mills theory, we can expand the $\mathcal{\hat A}(z)_\mu$ with respect to $z$ as 
\be\label{ExpA}
\mathcal{\hat A}(z)_\mu= A^1_\mu z +A^0_\mu+A^{-1_a}_\mu {1\over z-a}+A^{-1_b}_\mu {1\over z-b}+\cdots
\ee
According to (\ref{recz}),  the term $A^0_\mu$ should not contribute to the amplitude.  
Comparing (\ref{recz}) with (\ref{ExpA}), it is clean that the  boundary term should be 
\be\label{boundequ}
A^0\cdot \eta_i=-{A^1\cdot p_v } .
\ee
Therefore $A^0$, which is hard to obtain directly, can be transformed to the $A^1$ which can  be obtained  by a new recursion relation.

\section{Off-shell vector current from the Ward Identity}\label{WardinOff}
Now we apply our technology to the off-shell vector currents. Without loss of generality, we choose the shifted on-shell line to be of $+$ helicity. Under the gauge described  above, the overall behavior of the currents is $z^1$ when $z\rightarrow \infty$. 
According to eq. (\ref{recz}), we get
\bea\label{OffAz}
\eta^\mu_i \hat A_\mu&=& -\hat p_v^\mu A^1_\mu + \sum_m\sum_h \hat p_v^\mu {A_L^h(z_m)(A_R^{\td h})_\mu(z_m)\over 2P_m \cdot \eta_i (z-z_m)^2} \nb\\
&=& -\hat p_v^\mu A^1_\mu + \sum_m\sum_h  {A_L^h(z_m)(A_R^{\td h})_\mu(z_m) \eta^\mu_i \over 2P_m \cdot \eta_i (z-z_m)},
\eea
where the we choose the off-shell line in $A_R^{\td h}$.
Since the factor $A_L^{k_m}(z_m)$ vanishes, we only need to take the summation over $(h,\td h)\in \{(+,-),(-,+),(r,k)\}$\cite{Feng}. Then the current's projection on $\eta_i$ can be obtained by setting $z=0$ in (\ref{OffAz})
\bea\label{OffA}
\eta^\mu_i A_\mu
&=& -p_v^\mu A^1_\mu + \sum_m\sum_h  {A_L^h(z_m)(A_R^{\td h})_\mu(z_m)\eta^\mu_i \over 2P_m\cdot  \eta_i (-z_m)}.
\eea

Using the complexified Ward Identity, we can hence transform the unknown term $A^0$ into $A^1$ which can also be obtained by a new recursion relation.  In fact when we choose the momentum shift and the gauge of the external on-shell states as discussed above, the wave function does not depend on $z$. Moreover, the four point vertices do not contain the momentum factor,  they are also independent of  $z$. The $z$ dependent terms only come from the three point vertices and propagators in the complex lines from external line $i$ to $v$.  $A^1_\mu =({d\mathcal{\hat A}(z)_\mu \over dz})^0$ then gets two kinds of contributions as shown in Fig. 3.   
\begin{figure}[htb]
\centering
\includegraphics{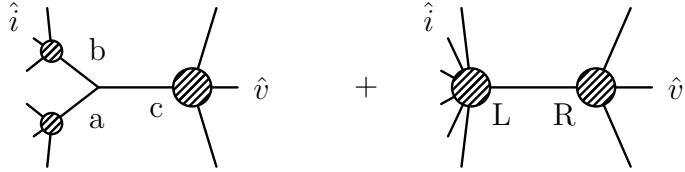}
\caption{The Feynman diagrams which will contribute to the boundary terms}
\end{figure}
When acting  with ${d\over dz}$ on the propagators and extracting the zeroth order terms in $z$,   we get 
\be
i( A_L^{\nu_L})^1{g_{\nu_L\nu_R}\over 2P_m\cdot\eta_i } (A_R^{\nu_R \mu})^1. 
\ee 
When acting with ${d\over dz}$ on the three point vertices, we get
\be
 {1\over \sqrt{2}} {A_a\cdot (A_b)^1~ \eta_i\cdot (A_c^\mu)^1\over 2p_b\cdot\eta_i~ 2p_c\cdot \eta_i ~p_a^2}-\sqrt{2} {(A_b)^1 \cdot(A_c^\mu)^1~ \eta_i\cdot A_a\over 2p_b\cdot\eta_i~ 2p_c\cdot \eta_i ~p_a^2}. 
 \ee
 Finally, we take the summation of all the complexified three-point vertices and the propagators 
\be
(A^{\mu})^1=\sum_{a,b,c}\left({1\over \sqrt{2}} {A_a\cdot (A_b)^1~ \eta_i\cdot (A_c^\mu)^1\over 2p_b\cdot\eta_i~ 2p_c\cdot \eta_i ~p_a^2} -\sqrt{2} {(A_b)^1 \cdot (A_c^\mu)^1~ \eta_i\cdot A_a\over 2p_b\cdot\eta_i~ 2p_c\cdot \eta_i ~p_a^2}\right)+ \sum_m i{(A_L)^1\cdot (A_R^\mu)^1\over 2p_m\cdot\eta_i}.
\ee

To complete the recursion relation, we need to know the coefficients of  order $z$  in the tensor off-shell  currents $(\hat A^{\nu\mu})^1$.  It is similar to the $(A^\mu)^1$,
\bea
(A^{\nu\mu})^1&=&{1\over \sqrt{2}} {A_a\cdot (A_b^\nu)^1~ \eta_i\cdot (A_c^\mu)^1\over 2p_b\cdot\eta_i~ 2p_c\cdot \eta_i ~p_a^2} -\sqrt{2} {(A_b^\nu)^1 \cdot (A_c^\mu)^1~ \eta_i\cdot A_a\over 2p_b\cdot\eta_i~ 2p_c\cdot \eta_i ~p_a^2}
+ i{(A_L^\nu)^1\cdot (A_R^\mu)^1\over 2p_m\cdot\eta_i}.
\eea

In (\ref{OffA}), there are new non-vanishing  objects which can be taken as the off-shell amplitudes with one external states of the on-shell lines replaced by its momentum.  On proceeding several recursive steps,  we get  a general form $\mathcal{A}^{\mu}(\cdots, k_{i_1},\cdots, k_{i_j},\cdots,k_{i_N})$, where we omit on-shell states with $N$ denotes the total number of the replaced on-shell lines.   
\begin{figure}[htb]
\centering
\includegraphics{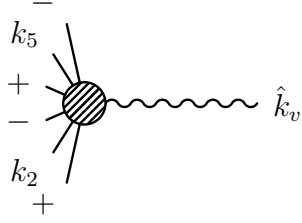}
\caption{Color-ordered Feynman rules.}
\end{figure}

Inevitably, we will need to shift the momentum of such line together with the off-shell line.  The boundary term then can not be obtained as discussed above.  Under the momentum  shifting,  $\lmd_i\rightarrow\lmd_i-z\lmd_v, \td\lmd_v\rightarrow\td\lmd_v+z\td\lmd_i$, the vector currents are of the form $\hat k^\nu_i \hat A_{\nu\mu}$.  The shifted momentum contains a factor proportional  to $z$. This will lead to $(\hat A^\mu)^0$ contributing to the boundary term. Luckily we only need to know the amplitude when $z\rightarrow 0$.  As it is obvious that $\hat k^\nu_i \hat A_{\nu\mu}|_{z=0}= k^\nu_i \hat A_{\nu\mu}|_{z=0}$, we only need to consider  $ k^\nu_i \hat A_{\nu\mu}|_{z=0}$.

There is a similar identity for calculating this kind of currents
\be\label{newId}
{ k^\nu_i \hat A_{\nu\mu} \hat K^\mu\over [\td\lmd_m,\td\lmd_v]}=0,
\ee
which can be deduced from the Ward identity, 
\bea
{\hat  k^\nu_i \hat A_{\nu\mu} \hat K^\mu \over [\td\lmd_m,\td\lmd_v]}&=&0,\nb\\
{\eta^\nu_i \hat A_{\nu\mu} \hat K^\mu\over [\td\lmd_m,\td\lmd_v]}&=&\epsilon^\nu_i \hat A_{\nu\mu} \hat K^\mu=0.
\eea

From (\ref{newId}), we get a similar recursion relation for $k^\nu_i  A_{\nu\mu} \eta_i^\mu$.  Similarly, for another momentum shift $\td\lmd_i\rightarrow\td\lmd_i-z\td\lmd_v, \lmd_v\rightarrow\lmd_v+z\lmd_i$, we get a recursion relation for $k^\nu_i  A_{\nu\mu} \td\eta_i^\mu$. 
Now we have successfully expressed  the boundary term in the amplitude as terms composed of  the off-shell amplitudes with fewer external lines. The component of the off-shell vector current in the momentum direction vanishes according to the Ward identity.  To get the full off-shell vector current, we hence need to project onto three linearly independent directions, non of which is  parallel with the momentum.   We can obtain all of them using the same procedure  above. 

Without loss of generality, we choose the helicities of lines $(i,j,k)$ to be $(+,+,-)$ respectively.  In the same way, we can obtain $\eta_j\cdot \ A$ and $\td\eta_k \cdot A$ where $\eta_j=\lmd_v\td\lmd_j$ and $\td\eta_k=\lmd_k\td\lmd_v$. It is convenient to write the vector current as $A^\mu=x_1\eta^\mu_i+x_2 \eta^\mu_j +x_3 \td\eta^\mu_k+x_4 K^\mu_v$. We can then determine the off-shell vector currents by solving the following four equations
\bea
e_{ik}x_3+e_{iv}x_4&=&\eta_i\cdot A\nb\\
e_{jk}x_3+e_{jv}x_4&=&\eta_j\cdot A\nb\\
e_{ki}x_1+e_{kj}x_2+e_{kv}x_4&=&\eta_k\cdot A\nb\\
e_{vi}x_1+e_{vj}x_2+e_{vk}x_3+e_{vv}x_4&=&0,
\eea
where $e_{ij}$ are the inner products of the basis vectors and the lower indices denote the corresponding basis. 
Hence to get the full vector current $A^\mu$, we only need to choose three different kinds of momentum shift, by choosing  various  external lines or shifts, such that the shifted momenta are linearly independent.  This is possible  for any vector currents with three external on-shell lines.

\section{Examples}
Before proceeding on the concrete examples, we summarize the procedure of calculating the vector currents. When the helicities of on-shell lines in the currents are same, it is direct to write the expressions of the currents as in \cite{Berends:1987me,Dixon1}. For the case when the external on-shell lines are of mixed helicity structures, we choose the reference spinors to be $\lmd_v$ and $\td\lmd_v$ for the $+$ and $-$ states respectively. We then choose three kinds of momentum shift such that all the external states are independent on the shifting parameter $z$ and the shifting momenta are linear independent with each other and not parallel with the momentum of the external off-shell line.  Under each kind of momenta shift, we then get one component of the current. Combining with the Ward identity, we can recover the whole vector current.  Each non-vanishing component of the complexified current is composed of a boundary term and single pole terms. For single pole terms, we get them by the BCFW recursion relation.  The referent vector, which will appear in cutting the internal propagators, are chosen to be $\lmd_v\td\lmd_v$ for the simplification of calculations.  For the boundary term, we get it by the new recursion relation in Section \ref{WardinOff}.

\subsection{currents with two on-shell lines}
There are two building blocks for  the recursion relations.   One is the three-point on-shell amplitude.  The form can be find  in  earlier papers \cite{Berends:1987me,Badger1}.  The other is the off-shell current with three external lines with one off-shell external line and two on-shell external lines.  The external vector of every on-shell line  can be physical state or its momentum.  The off-shell currents take the following forms,
\begin{figure}[htb]
\centering
\includegraphics{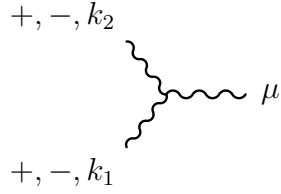}
\caption{Color-ordered Feynman rules.}
\end{figure}
\bea
A(k_1,k_2,\mu)&=&{i\over \sqrt{2}}k_1\cdot k_2 (k_2-k_1)_{\mu}\nb\\
A(\pm,k_2,\mu)&=&{i\over \sqrt{2}}\left(k_2\cdot \epsilon^{\pm}_1 (k_2+k_1)_{\mu}-2k_1\cdot k_2 (\epsilon^{\pm}_1)_\mu \right)\nb\\
A(k_1,\pm,\mu)&=&{i\over \sqrt{2}}\left(-k_1\cdot \epsilon^{\pm}_2 (k_2+k_1)_{\mu}+2k_1\cdot k_2 (\epsilon^{\pm}_2)_\mu \right)
\eea
\bea
A(\pm,\pm,\mu)&=&{i\over \sqrt{2}}\left(2k_2\cdot \epsilon^{\pm}_1 (\epsilon^{\pm}_2)_{\mu}-2k_1\cdot \epsilon^{\pm}_2 (\epsilon^{\pm}_1)_\mu \right)\nb\\
A(\mp,\pm,\mu)&=&{i\over \sqrt{2}}\left(\epsilon^{\mp}_1\cdot \epsilon^{\pm}_2 (k_1-k_2)_{\mu}+2k_2\cdot \epsilon^{\mp}_1 (\epsilon^{\pm}_2)_{\mu}-2k_1\cdot \epsilon^{\pm}_2 (\epsilon^{\mp}_1)_\mu \right).
\eea
There is a direct justification of equation (\ref{boundequ}) for three-line off-shell amplitudes.  We take $A(+,-,\mu)$ as an example. We choose the $+$-momentum shift for lines $1$ and $3$. The reference spinors of the external lines are the same as above.  It is easy to show that  $A^0\cdot \eta_1=-A^1\cdot k_3$:
\bea
A^0\cdot \eta_1&=&{i\over \sqrt{2}}\left(-\epsilon^{+}_1\cdot \epsilon^{-}_2 k_2\cdot \eta_1+2k_2\cdot \epsilon^{+}_1 \epsilon^{-}_2\cdot\eta_1\right) ={i\over \sqrt{2}} \epsilon^{+}_1\cdot \epsilon^{-}_2 k_2\cdot \eta_1\nb\\
-A^1\cdot k_3&=&{i\over \sqrt{2}}\left(\epsilon^{+}_1\cdot \epsilon^{-}_2 k_3\cdot \eta_1-2k_3\cdot \epsilon^{+}_1 \epsilon^{-}_2\cdot\eta_1 \right)={i\over \sqrt{2}} \epsilon^{+}_1\cdot \epsilon^{-}_2 k_2\cdot \eta_1.
\eea

\subsection{currents with three on-shell lines}
The first non-trivial example is an amplitude with four lines, one of which is off-shell. For concreteness,  we take the amplitude as $A(1^+,2^+,3^-, \mu_4)$.  As stated above, the reference momenta and spinors are taken as $k_r=\lmd_r\td\lmd_r$ and $\mu_1=\mu_2=\lmd_r, \td\mu_3=\td\lmd_r$.  The shifted momenta  are $\eta_1=\lmd_r\td\lmd_1,\eta_2=\lmd_r\td\lmd_2,\td\eta_3=\td\lmd_r \lmd_3$ for  the states $\epsilon^{+}_1={\lmd_r\td\lmd_1 \over \la\lmd_r,\lmd_1\ra},\epsilon^{+}_2={\lmd_r\td\lmd_2 \over \la\lmd_r,\lmd_2\ra},\epsilon^{-}_3={\lmd_3\td\lmd_r \over [\td\lmd_3,\td\lmd_r]}$ respectively. Through these shifts, one can get the  components $A\cdot \eta_1, A\cdot \eta_2, A \cdot  \td\eta_3$ of the off-shell amplitude vector.  We will compare the results from our methods with those from the usual Feynman rules for $A\cdot \eta_1$.  Similar discussions are valid for  $A\cdot \eta_2, A \cdot  \td\eta_3$.  In Feynman rules the four-line amplitude involves the following three diagrams. 
\begin{figure}[htb]
\centering
\includegraphics{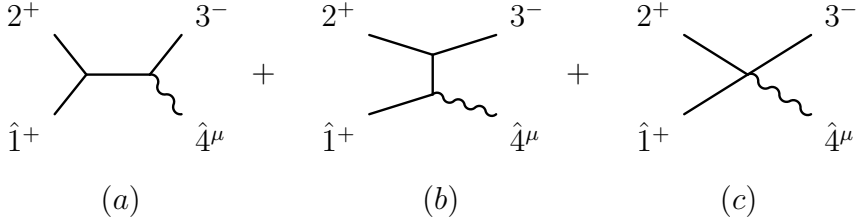}
\caption{Color-ordered Feynman rules.}
\end{figure}
Figure (a) contributes a term 
\bea\label{aterm}
{A(1^+, 2^+,\nu_1)  g^{\nu_1\nu_2}A(\nu_2,3^-,\mu)\eta^\mu \over p_{12}^2}.
\eea
Figure (b) is
\bea\label{bterm}
&&{1\over \sqrt 2}{\epsilon_1\cdot A(\mu,2^+,3^-) (k_1-k_{14})\cdot\eta_1+2\eta_1\cdot A(\mu,2^+,3^-) k_{14}\cdot \epsilon_1\over p_{23}^2}\nb\\
&=&{1\over \sqrt 2}{\eta_1\cdot A(\mu,2^+,3^-)(k_1+ k_{14})\cdot \epsilon_1\over p_{23}^2}\nb\\
&=&{-1\over \sqrt 2}{\eta_1\cdot A(\mu,2^+,3^-) k_{4}\cdot \epsilon_1\over p_{23}^2}.
\eea
Figure (c) vanishes for $A\cdot \eta_1$. 
Using our methods, we  get the off-shell amplitude as 
\bea\label{ex1}
A\cdot \eta_1&=&-p_4\cdot A^1+{A(\hat 1^+, 2^+,\hat m^-) A(\hat m^+,3^-,\hat 4^\mu)\eta^\mu \over2 P_m\cdot \eta (-z_m)}\nb\\
&=&-{1\over \sqrt 2}{\eta\cdot A(\mu,2^+,3^-) \epsilon_{\hat 1^+}\cdot p_4\over p_{23}^2}+{A(\hat 1^+, 2^+,\hat m^-) A(\hat m^+,3^-,\hat 4^\mu)\eta^\mu \over p_{12}^2}.
\eea
The first term is equal to (\ref{bterm}). The second term  can be rewritten as 
\bea
&&{A^{z_m}(\hat 1^+, 2^+,\hat m^-) A^{z_m}(\hat m^+,3^-,\hat 4^\mu)\eta^\mu \over p_{12}^2}={A^{z_m}(\hat 1^+, 2^+,\nu_1) g^{\nu_1\nu_2}A^{z_m}(\nu_2,3^-,\hat 4^\mu)\eta^\mu \over p_{12}^2}\nb\\
&=&{\left(A(1^+, 2^+,\nu_1) +{i\over \sqrt{2}}\left(z_m\epsilon^{+}_1\cdot \epsilon^{+}_2 (-\eta)_{\nu_1}+2z_m\eta_1\cdot \epsilon^{+}_2 (\epsilon^{+}_1)_{\nu_1}\right)\right) g^{\nu_1\nu_2}A^{z_m}(\nu_2,3^-,\hat 4^\mu)\eta^\mu \over p_{12}^2}\nb\\
&=&{A(1^+, 2^+,\nu_1)  g^{\nu_1\nu_2}\left(A(\nu_2,3^-, 4^\mu)\eta_1^\mu+{i\over \sqrt{2}}\left(z_m (\epsilon^{+}_2)_{v_2} (-\eta_1)_{\mu}\eta_1^\mu+2z_m (\eta_1)_{v2}\eta_1\cdot \epsilon^{-}_3 \right)\right) \over p_{12}^2}\nb\\
&=&{A(1^+, 2^+,\nu_1)  g^{\nu_1\nu_2}A(\nu_2,3^-,\mu)\eta^\mu \over p_{12}^2},\nb\\
\eea
which is exactly the same as  (\ref{aterm}). 
When shifting the momenta of a pair of lines $(2, 4)$,  we get 
\bea
\eta_2\cdot A&=&-p_4 \cdot A^1+{A(1^+,\hat 2^+, \hat m_{12}^-) A(\hat m_{34}^+, 3^-, \hat 4^\mu) \eta_2^\mu \over 2p_{34}\cdot\eta_2 (-z_{m_{12}})}\nb\\
&+&{A(\hat 2^+, 3^-, \hat m_{23}^+) A(\hat m_{14}^-, \hat 4^\mu,1^+) \eta_2^\mu \over 2p_{14}\cdot\eta_2 (-z_{m_{12}})}+{A(\hat 2^+, 3^-, \hat m_{23}^v) A(\hat m_{14}^k, \hat 4^\mu,1^+)\cdot \eta_2^\mu \over 2p_{14}\cdot\eta_2 (-z_{m_{12}})},
\eea
where 
\be
-p_4 \cdot A^1={1\over \sqrt 2}{\eta_2^\mu A^1(m_{14}^\mu, \hat 4^\nu, 1^+) p_4^\nu \epsilon^+_2\cdot \epsilon_3^- \over 2p_{14}\cdot\eta_2 (-z_{m_{12}})}-i{A^1(\hat 2^+, 3^-, m_{23}^\mu) A^1(m_{14}^\mu, \hat 4^\nu,1^+)p_4^\nu \over 2p_{14}\cdot\eta_2 }
\ee
For shifting the momenta the pair of lines $(3,4)$, we get 
\bea
\td\eta_3\cdot A&=&-p_4 \cdot A^1+{A(2^+,\hat 3^-, \hat m_{23}^-) A(\hat m_{14}^+,\hat 4^\mu, 1^+) \eta_3^\mu \over 2p_{14}\cdot\td\eta_3 (-z_{m_{14}})}\nb\\
&+&{A(2^+,\hat 3^-, \hat m_{23}^+) A(\hat m_{14}^-,\hat 4^\mu, 1^+) \eta_3^\mu \over 2p_{14}\cdot\td\eta_3 (-z_{m_{14}})}
+{A(2^+,\hat 3^-, \hat m_{23}^r) A(\hat m_{14}^k,\hat 4^\mu, 1^+) \eta_3^\mu \over 2p_{14}\cdot\td\eta_3 (-z_{m_{14}})},
\eea
where 
\bea
-p_4\cdot A^1&=&{1\over \sqrt 2}{\td\eta_3\cdot p_4 A(1^+, 2^+, \mu)\epsilon^\mu_{3^-} \over p^2_{12}}
+{1\over \sqrt 2}{\td\eta_3^\mu A^1(m_{14}^\mu, \hat 4^\nu, 1^+) p_4^\nu \epsilon^+_2\cdot \epsilon_3^- \over 2p_{14}\cdot\td\eta_3}\nb\\
&+&{1\over \sqrt 2}{\td\eta_3\cdot p_4 A^1(2^+, \hat 3^+, m_{23}^\mu)\epsilon^\mu_{1^+} \over  2p_{14}\cdot\td\eta_3}
-i{A^1(2^+, \hat 3^-, m_{23}^\mu) A^1(m_{14}^\mu, \hat 4^\nu,1^+)p_4^\nu \over 2p_{14}\cdot\td\eta_3}.
\eea
\subsection{currents with four on-shell lines}
Now we apply our techniques to the currents with four on-shell external line $\mathcal{A}(1^+, 2^+, 3^+, 4^-, \mu)$. To get the full components of the currents, we choose three shifted pairs of lines $(1,5), (2,5), (4,5)$. 

For the shifted pair $(1,5)$, the finite poles contributions to the currents come from the Feynman diagrams in Fig. 7. The infinite contributions come from the Feynman diagrams in Fig.8.
\begin{figure}[htb]
\centering
\includegraphics{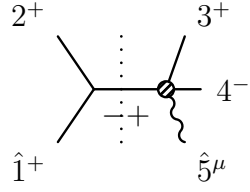}
\caption{Finite pole terms in the momentum shift of $(1,5)$}
\end{figure}
\begin{figure}[htb]
\centering
\includegraphics{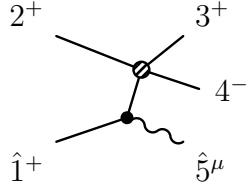}
\caption{Boundary terms in the momentum shift of $(1,5)$. Here we use the $\cdot$ to denote the new vertex which is the result of acting $d\over dz$ on the original three point vertex.}
\end{figure}
Then we have
\bea
\eta_1\cdot A&=&-p_5 \cdot A^1+{A(\hat 1^+,2^+, \hat m_{12}^-) A(\hat m_{345}^+,3^+, 4^-,\hat 5^\mu) \eta_1^\mu \over 2p_{345}\cdot\eta_1 (-z_{m_{12}})},
\eea
where 
\bea
-p_5 \cdot A^1={1\over \sqrt 2}{ \eta_1^\mu A(m_{234}^\mu,2^+, 3^+, 4^-) \epsilon^{\hat 1^+}\cdot  p_5 \over p^2_{234}}.
\eea
For the momenta shift of the pair  (2,5), we have 
\bea
\eta_2\cdot A&=&-p_5 \cdot A^1-i{A(1^+, \hat 2^+, \hat m_{12}^-) A(\hat m_{345}^+,3^+, 4^-,\hat 5^\mu) \eta_2^\mu \over 2p_{345}\cdot\eta_2 (-z_{m_{12}})}\nb\\
&-& i{A(\hat 2^+,3^+, \hat m_{23}^-) A(\hat m_{451}^+,4^-,\hat 5^\mu,1^+) \eta_2^\mu \over 2p_{451}\cdot\eta_2 (-z_{m_{23}})}\nb\\
&-& i{A(\hat 2^+,3^+, 4^-, \hat m_{234}^-) A(\hat m_{51}^+,\hat 5^\mu,1^+) \eta_2^\mu \over 2p_{51}\cdot\eta_2 (-z_{m_{234}})}\nb\\
&-& i{A(\hat 2^+,3^+, 4^-, v) A(\hat k,\hat 5^\mu,1^+) \eta_2^\mu \over   2p_{51}\cdot\eta_2 (-z_{m_{234}}) \hat k\cdot v },\nb\\
\eea
where 
\bea
-p_5 \cdot A^1&=&\sqrt 2{ \eta_2\cdot \epsilon_{1}^+  A^1(m_{234}^\mu,\hat 2^+, 3^+, 4^-) p^\mu_5 \over 2 p_{15}\cdot \eta_2}
-{1\over \sqrt 2}{ \eta_2\cdot p_5  A^1(m_{234}^\mu,\hat 2^+, 3^+, 4^-) \epsilon_{ 1}^+ \over 2 p_{15}\cdot \eta_2}\nb\\
&-&{1\over \sqrt 2}{ \eta_2^\mu  A(m_{34}^\mu, 3^+, 4^-) \epsilon_{{\hat 2}^+}^{\mu_2}A^1(m_{15}^{\mu_2}, \hat 5^{\mu_1},1)p^{\mu_1} \over 2 p_{15}\cdot \eta_2 p_{34}^2}\nb\\
&-&i{ A^1(m_{234}^\mu,\hat 2^+, 3^+, 4^-)  A^1(m_{51}^\mu, \hat 5^\nu,1^+) p^\nu \over 2p_{15}\cdot \eta_2}.
\eea
The diagrams contributing to the finite pole terms and boundary terms are shown in Fig.9 and Fig.10 respectively. 
\begin{figure}[htb]
\centering
\includegraphics{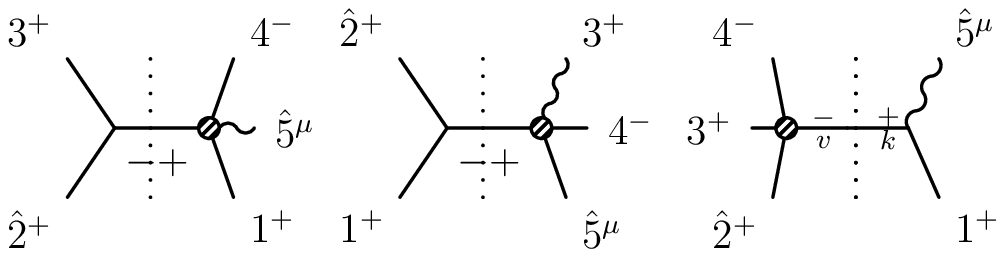}
\caption{Finite pole terms in the momentum shift of $(2,5)$}
\end{figure}
\begin{figure}[htb]
\centering
\includegraphics{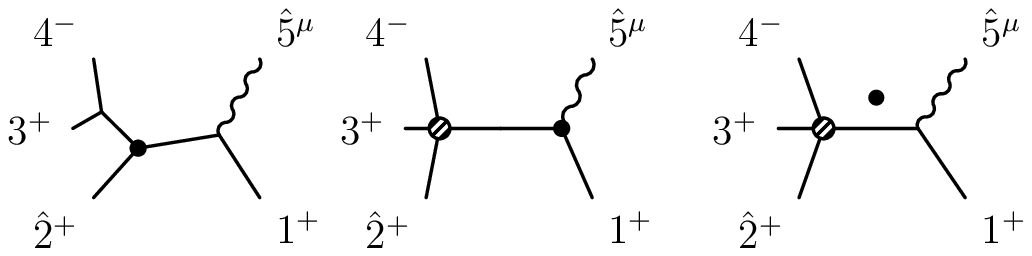}
\caption{Boundary terms in the momentum shift of $(2,5)$. Here we use the $\cdot$ to denote that we should act  the $d\over dz$ on the corresponding three point vertex or propagator first.}
\end{figure}
For the momentum shift of the pair (4,5), we have
\bea
\td\eta_3\cdot A&=&-p_5 \cdot A^1-i{\td\eta_3^\mu A(\hat 5^\mu, 1^+, \hat m_{15}^{\substack{+\\k}}) A(\hat m_{234}^{\substack{-\\v}},2^+, 3^+, \hat 4^-)  \over 2p_{15}\cdot\eta_2 (-z_{m_{15}})}\nb\\
&-& i{\td\eta_3^\mu A(\hat 5^\mu, 1^+, 2^+, \hat m_{125}^{\substack{+\\-\\k}}) A(\hat m_{34}^{\substack{-\\+\\v}},3^+, \hat 4^-)  \over 2p_{125}\cdot\eta_2 (-z_{m_{125}})}\nb\\
\eea
\begin{figure}[htb]
\centering
\includegraphics{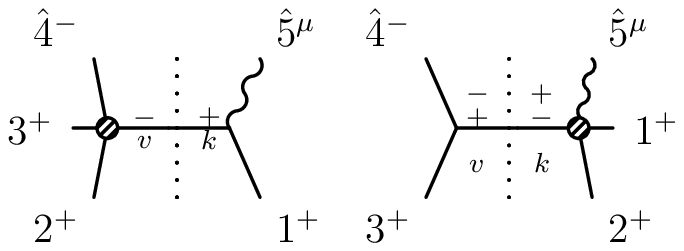}
\caption{Finite pole terms in the momentum shift of $(4,5)$}
\end{figure}
and the boundary term is 
\bea
&&-p_5 \cdot A^1=\sqrt 2{ \td\eta_3\cdot \epsilon_{1}^+  A^1(m_{234}^\mu, 2^+, 3^+, \hat 4^-) p^\mu_5 \over 2 p_{15}\cdot \td\eta_3}
-{1\over \sqrt 2}{ \td\eta_3\cdot p_5  A^1(m_{234}^\mu,\hat 2^+, 3^+, \hat 4^-) \epsilon_{ 1}^+ \over 2 p_{15}\cdot \td\eta_3}\nb\\
&+&\sqrt 2{ \td\eta^\mu_3\cdot A(m_{12}^\mu,1^+,2^+)  A^1(m_{34}^\nu, 3^+, \hat 4^-) p^\nu_5 \over 2 p_{125}\cdot \td\eta_3~ p^2_{12}} 
-{1\over \sqrt 2}{ \td\eta_3\cdot p_5  A^1(m_{34}^\mu, 3^+, \hat 4^-) A(m_{12}^\mu,1^+,2^+) \over 2 p_{125}\cdot \td\eta_3 ~p^2_{12}} \nb\\
&+&\sqrt 2{ \td\eta^\mu_3\cdot A(m_{123}^\mu,1^+,2^+,3^+)  \epsilon_{\hat 4^-} \cdot p_5 \over ~ p^2_{123}} 
-{1\over \sqrt 2}{ \td\eta_3\cdot p_5  \epsilon^\mu_{\hat 4^-}A(m_{123}^\mu,1^+,2^+,3^+) \over p^2_{123}} \nb\\
&+&\sqrt 2{ \td\eta^\mu_3\cdot \epsilon_{2^+}  A^1(3^+,\hat 4^-,m_{34}^{\mu_1})   A^1(m_{15}^{\mu_1}, \hat 5^{\mu_2},1^+)  p^{\mu_2}_5 \over ~ 2p_{125}\cdot \td\eta_3 ~ 2p_{15}\cdot \td\eta_3} 
-{1\over \sqrt 2}{ \td\eta^{\mu_1}_3 A^1(m_{15}^{\mu_1}, \hat 5^{\mu_2},1^+)  p^{\mu_2}_5   \epsilon^\mu_{2^+}  A^1(3^+,\hat 4^-,m_{34}^{\mu})    \over ~ 2p_{125}\cdot \td\eta_3 ~ 2p_{15}\cdot \td\eta_3}  \nb\\
&+&\sqrt 2{ \td\eta^\mu_3 A(2^+, 3^+,m_{23}^{\mu})    \epsilon_4^{\mu_1} A^1(m_{15}^{\mu_1}, \hat 5^{\mu_2},1^+)  p^{\mu_2}_5 \over ~ 2p_{23}^2~ 2p_{15}\cdot \td\eta_3} 
-{1\over \sqrt 2}{ \td\eta^{\mu_1}_3  A^1(m_{15}^{\mu_1}, \hat 5^{\mu_2},1^+)  p^{\mu_2}_5 ~A(2^+, 3^+,m_{23}^{\mu})    \epsilon_4^{\mu} \over ~ 2p_{23}^2~ 2p_{15}\cdot \td\eta_3}  \nb\\
&+&\sqrt 2{ \td\eta^\mu_3\cdot \epsilon_{3^+}   \epsilon_{\hat 4^-}^{\mu_1} A^1(m_{512}^{\mu_1}, \hat 5^{\mu_2},1^+, 2^+)  p^{\mu_2}_5 \over ~ 2p_{125}\cdot \td\eta_3 } 
-{1\over \sqrt 2}{ \td\eta^{\mu_1}_3  A^1(m_{512}^{\mu_1}, \hat 5^{\mu_2},1^+, 2^+)  p^{\mu_2}_5   \epsilon_{3^+} \cdot \epsilon_{4^-} \over ~ 2p_{125}\cdot \td\eta_3 }  \nb\\
&-&i{ A^1(m_{234}^\mu, 2^+, 3^+, \hat 4^-)  A^1(m_{51}^\mu, \hat 5^\nu,1^+) p_5^\nu \over 2p_{15}\cdot \td\eta_3}\nb\\
&-&i{ A^1(m_{34}^\mu, 3^+, \hat 4^-)  A^1(m_{512}^\mu,\hat 5^\nu, 1^+, 2^+) p_5^\nu \over 2p_{15}\cdot \td\eta_3}.
\eea
The corresponding diagrams of the finite pole terms and boundary terms are shown in Fig.11, Fig.12, Fig.13 respectively. 
\begin{figure}[htb]
\centering
\includegraphics{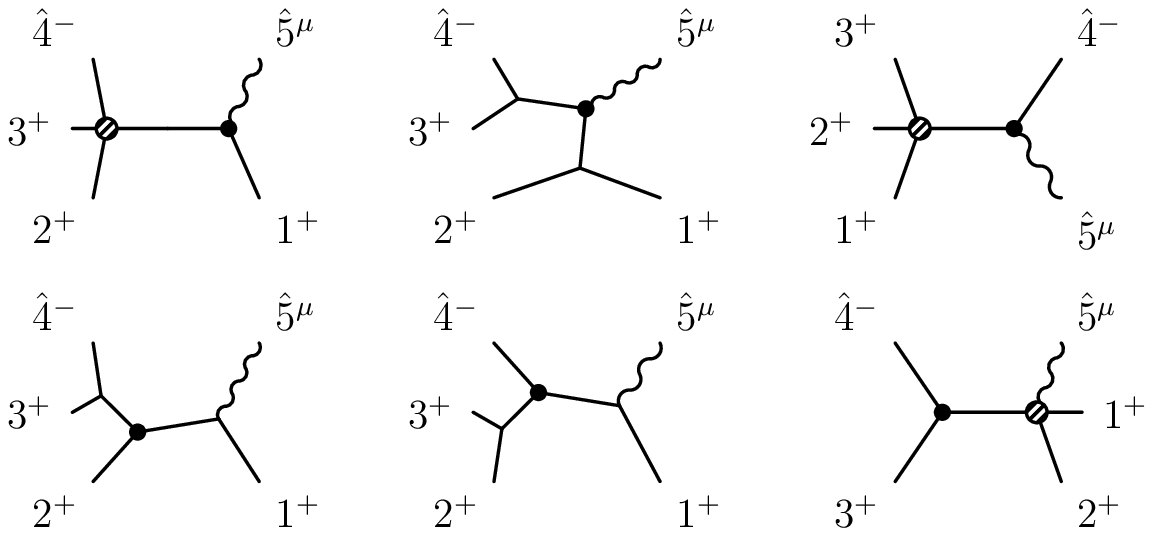}
\caption{Boundary terms in the momentum shift of $(4,5)$. Here we use the $\cdot$ to denote that we should act  the $d\over dz$ on the corresponding three point vertex  first.}
\end{figure}
\begin{figure}[htb]
\centering
\includegraphics{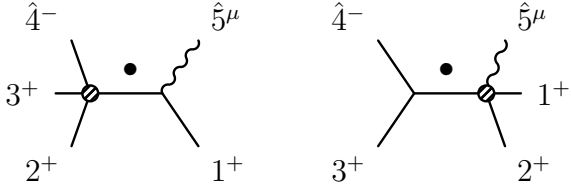}
\caption{Boundary terms in the momentum shift of $(4,5)$. Here we use the $\cdot$ to denote that we should act  the $d\over dz$ on the corresponding  propagator first.}
\end{figure}

\section{conclusion}
In this article, we analysis the cancelation in detail among the terms in Ward identity at tree-level in Feynman gauge. According to this, we prove the complexified Ward identity for the tree-level amplitudes directly.   Using the complexified Ward identity, we get exactly the usual BCFW recursion relation when reduced to single poles. Furthermore, we propose a new form for the boundary term for the BCFW shifted amplitude or off-shell vector currents. In this way, the boundary terms can be obtained by a new recursion relation.  We apply our technique to the off-shell currents with on-shell lines  with different helicity structures. It is easy to see that our technique is more efficient for the currents of general helicity structures for the on-shell lines, complimenting the existent off-shell recursion relation. First of all, the number of the effective diagrams is small.  For finite poles contribution, only the propagator along the complex line contributes; while for the  boundary terms  both the propagator and half parts in the three-point vertex  contribute to the vector currents. Four-point gluon interaction needs no consideration.  In proceeding the recursion relation, there will be new off-shell currents with some of the on-shell states  replaced by their momenta.  Such new objects can also be obtained in our techniques.

Although we focus on the one-line off-shell vector currents in gauge theory, the technique from complex Ward identity can be generalize to theories with gauge symmetry spontaneously broken as well as to tensor currents with several off-shell lines.  The current with two off-shell line is important for constructing one-loop amplitudes. Another extension is to study the amplitude at one loop level according to the loop level Ward identity. However,  the complex Ward identity does not present itself at the loop level, it warrants further study.  

\section*{Acknowledgement}
We thank Edna Cheung, Jens Fjelstad, Konstantin G. Savvidy and Yun Zhang for helpful discussions.  This work is funded by the Priority Academic Program Development of Jiangsu Higher Education Institutions (PAPD), NSFC grant No.~10775067, Research Links Programme of Swedish Research Council under contract No.~348-2008-6049, the Chinese Central Government's 985 Project grants for Nanjing University, the China Science Postdoc grant no. 020400383. the postdoc grants of Nanjing University 0201003020

\end{document}